\documentstyle[psfig,12pt]{article}

\def\beq{\begin{equation}}
\def\eeq{\end{equation}}
\begin{document}

%----------------------------------------------------------------------
%                     T I T L E
%----------------------------------------------------------------------

\begin{titlepage}

\begin{flushright}
WIS99/27/Aug-DPP

hep-th/0005075
\end{flushright}

\begin{centering}

\vspace*{3cm}

{\Large\bf On the Bosonic Spectrum of QCD(1+1) with SU(N) Currents}

\vspace*{1.5cm}

{\bf Uwe Trittmann}
\vspace*{0.5cm}

%{$^b$
{\sl Department of Physics\\
Ohio State University\\
Columbus, OH 43210, USA}

\vspace*{1cm}

%{Version, 08 X 2001}
%\vspace*{1cm}

%----------------------------------------------------------------------%
%                       A B S T R A C T
%----------------------------------------------------------------------%
\vspace*{2cm}

{\large Abstract}

\vspace*{1cm}

\end{centering}

In this note we calculate the spectrum of 
two-dimensional QCD. We formulate the theory with $SU(N_c)$ 
currents rather than 
with fermionic operators. We construct the Hamiltonian matrix in DLCQ 
formulation as a function of the harmonic resolution $K$ and the numbers 
of flavors $N_f$ and colors $N_c$. 
The resulting numerical eigenvalue spectrum is free from 
trivial multi-particle states which obscured previous results.
The well-known 't Hooft and large $N_f$ spectra are reproduced. In the case of 
adjoint fermions we present some new results.
\vspace{0.5cm}

%\noindent
%PACS number(s): 

\vfill
%$^*$ former address: {\em Department of Particle Physics, 
%Weizmann Institute of Science, 76100 Rehovot, Israel}

\end{titlepage}
\newpage

%----------------------------------------------------------------------%
%                       Introduction
%----------------------------------------------------------------------%

\section{Introduction}

\begin{figure}
\centerline{
\psfig{file=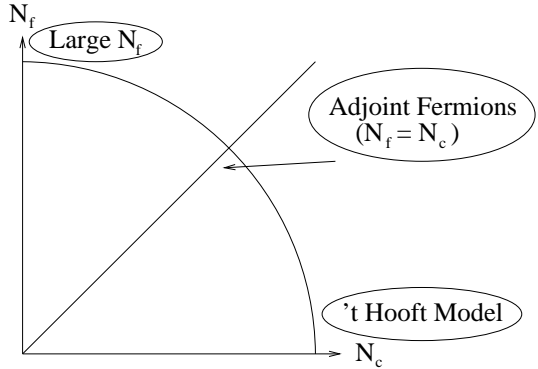,width=7.0true cm,angle=0}
}
\caption{The parameter space of two-dimensional Yang-Mills 
theories with massless fermions.}
\label{Fig0}
\end{figure}

On the way to understand the physics of strong interactions,  
two-dimensional QCD has remained an interesting model ever since it has
been studied by 't Hooft \cite{tHooft74b}. 
It has, however, the disadvantage of lacking dynamical 
({\em i.e.}~transverse) gluonic degrees of freedom.
To construct a model closer to four dimensional QCD, adjoint 
rather than fundamental fermions were 
built into the theory \cite{DalleyKlebanov93b}\cite{Kutasov94}. 
Also matter with a large number of flavors has been considered
\cite{DateFrishmanSonnenschein87}\cite{ArmoniSonnenschein95}. 
Surprisingly, these theories are related 
by a universality \cite{KutasovSchwimmer95}:
the massive spectrum and interactions of two-dimensional Yang-Mills theories
coupled to massless fermions in 
arbitrary representations depend only on the gauge group and the level 
of the affine Lie algebra.
The associated parameter space of two-dimensional Yang-Mills theories
with massless fermions is depicted in Fig.~\ref{Fig0} \cite{AFST98}.

Of particular interest in two-dimensional QCD is the transition between 
confinement and screening. The 't Hooft model is known to consist
of stable mesons and has no confinement-screening transition.
The model with adjoint fermions exhibits deconfinement
at zero fermion mass $m$, and
the string tension vanishes like $\sigma\propto m$ at this point 
\cite{Smilga96}\cite{FrishmanSonnenschein95}\cite{ArmoniFrishmanSonnenschein98}.
Interestingly, this theory is supersymmetric at the fermion mass 
$m^2=g^2N_c$, and asymptotically supersymmetric otherwise.
It has been studied extensively in the literature
\cite{DalleyKlebanov93b,Kutasov94,GHK97,AntonuccioPinsky98,BhanotDemeterfiKlebanov93}.
We will focus on the massless case of this theory, where we can make use 
of the universality 
\cite{KutasovSchwimmer95}, and consequently we will formulate the theory as a 
current algebra problem. The Hilbert space of the problem then splits up 
into current sectors.
This allows for a convenient reduction of the numerical effort when
solving for the spectrum of the theory, since 
all single-particle states lie in the current block of the identity (bosons)
or in the adjoint current block (fermions) \cite{KutasovSchwimmer95}. 
Because of its simpler algebraic properties, we will limit ourselves in
the present work to the bosonic spectrum.  
We will use the additional parameter in the problem, $\lambda=N_f/N_c$,
as a tool for 
interpreting the spectrum, {\em i.e.}~we will consider  
theories 'on the arc' of Fig.~\ref{Fig0}, where both $N_c$ and $N_f$ are
large. 
The main emphasis will be on the adjoint case ($N_c{=}N_f$).
The goal is to 
identify the single-particle states of the theories and to be able
to connect to results anticipated from 
the asymptotic theory, {\em e.g.}~the expected multi-Regge trajectory 
structure in the adjoint case \cite{Kutasov94}.

A very convenient way to formulate two-dimensional QCD is to quantize 
the system on the light-cone \cite{tHooft74b}.
Usually the light-cone gauge, $A^+{=}0$, is used. This approach gives a very
complete picture and includes non-perturbative effects, if an appropriate 
regularization for the quark self-energy is used \cite{Bassetto98}.  
Discretized Light Cone Quantization (DLCQ) 
\cite{PauliBrodsky85a} is then a method especially suited to solve 
numerically for the spectrum of low dimensional theories. 
The momenta are discretized by imposing boundary conditions on the fields.
The typical DLCQ program is to construct a finite dimensional 
Hamiltonian matrix characterized by the harmonic resolution $K$.
The spectrum is obtained by diagonalizing this matrix numerically
for larger and larger $K$, and eventually extrapolating to the continuum
limit, $K\rightarrow\infty$.

The latest large $N_c$ 
analysis of adjoint QCD$_2$ \cite{GHK97} 
revealed several single-particle states and, most 
interestingly, a continuum of states at precisely four times the mass (squared)
of the lowest fermion state. Despite this remarkable structure, the
insight gained from this finding is rather small. 
It is clearly not a signature for screening versus confinement. 
Both in a confining and a screening theory, one would
expect a continuum of states at $4m^2$ of a single-particle state, although 
from very different mechanisms. 
More information about the spectrum is thus needed to understand things
like the deconfinement mechanism or the statistics of single-particle
states.

The paper is organized as follows. In the following section we review 
two-dimensional QCD with massless fermions.
Sec.~\ref{SecAnalytic} deals with analytic considerations. Namely, we show
that it is possible to calculate two eigenvalues of the
theory from the outset. 
In Sec.~\ref{SecHamiltonian} we construct the Hamiltonian matrix 
for any harmonic resolution $K$ and
arbitrary ({\em i.e.}~in principle finite) $N_f$ and $N_c$. We will,
however, {\em in praxi} only use the large $N$ limit of this result. 
Sec.~\ref{SecNumCalc} details the numerical results. 
We will focus on the adjoint case, but also consider the 't Hooft limit and 
the large $N_f$ limit to present a coherent picture of two-dimensional QCD.
A discussion of the results follows.

%----------------------------------------------------------------------%
%                       QCD in two dimensions
%----------------------------------------------------------------------%

\section{QCD in two dimensions}

We want to compute the massive spectrum of $SU(N_c)$ Yang-Mills gauge 
fields coupled to massless adjoint fermions in two dimensions. Due to the 
universality in these gauge theories \cite{KutasovSchwimmer95}, this is 
equivalent to solving for the massive spectrum of $N_f{=}N_c$ flavors of
massless fundamental Dirac fermions coupled to the gauge fields. 
The latter case turns out to be more general, in the sense that we have 
an additional continuous parameter, $\lambda{:=}N_f/N_c$, in the theory.
The case of adjoint gauged fermions is recovered when setting
$\lambda$ to unity.
Without loss of generality we consider only the Lagrangian of the 
fundamental theory
\beq
{\cal L}=Tr[-\frac{1}{4g^2}F_{\mu\nu}F^{\mu\nu}+
i\sum_{a=1}^{N_f}\bar{\Psi}_a\gamma_{\mu}D^{\mu}\Psi_a]
\eeq
where $\Psi_{a}=2^{-1/4}({\psi_{a} \atop \chi_{a}})$, 
with $\psi_a$ and $\chi_a$ 
carrying color indices, which we suppressed. The field strength is
$F_{\mu\nu}=\partial_{\mu}A_{\nu}-\partial_{\nu}A_{\mu}+i[A_{\mu},A_{\nu}]$,
and the covariant derivative is defined as $D_{\mu}=\partial_{\mu}
+i[A_{\mu},\cdot]$. 
We work in the light-cone gauge $A^+=0$ and use the convenient Dirac basis
$\gamma^0=\sigma_1$, $\gamma^1=-i\sigma_2$. 
The Lagrangian then becomes
\beq
{\cal L}=Tr\left[\frac{1}{2g^2}(\partial_-A_+)^2+i\psi^{\dagger}\partial_
+\psi+i\chi^{\dagger}\partial_-\chi-A_+J\right],
\eeq
with the current $J_{ij}=\psi^{\dagger}_{ia}\psi_{aj}$.
We can integrate out the 
(non-dynamical) component $A_+$ of the gauge field and obtain 
\beq
{\cal L}=Tr\left[i\psi^{\dagger}\partial_+\psi+i\chi^{\dagger}\partial_-\chi
-\frac{g^2}{2} J\frac{1}{\partial^2_-}J\right].
\eeq
It is obvious that the left-movers $\chi$ decouple.
Noting the simple expression of the interaction in terms of the
currents, 
it is natural to bosonize the theory \cite{Witten84}.
We follow Ref.~{\cite{ArmoniSonnenschein95}} to derive the momentum operators.
The bosonized action of colored flavored fermions is
\beq
S_0=S^{WZW}_{(N_f)}(g)+S^{WZW}_{(N_c)}(h)+\frac{1}{2}\int d^2x\partial_{\mu}
\phi\partial^{\mu}\phi,
\eeq
where $g\in SU(N_c)$, $h\in SU(N_f)$,
$\exp\{i\sqrt{\frac{4\pi}{N_cN_f}}\phi\} \in U_B(1)$, and the 
Wess-Zumino-Witten action 
is
\beq
S^{WZW}_{(k)}(g)=\frac{k}{8\pi}\int_{\Sigma} d^2x 
Tr(\partial_{\mu}g\partial^{\mu}g^{-1})
+\frac{k}{12\pi}\int_{B} d^3y\epsilon^{ijk} 
Tr(g^{-1}\partial_{i}g)(g^{-1}\partial_{j}g)(g^{-1}\partial_{k}g),
\eeq
where $B$ is the solid sphere whose boundary $\Sigma$ represents space-time.
The action of the theory is then
\beq
S=S_0+\frac{g^2}{2}\int d^2x J\frac{1}{\partial^2_-}J, 
\eeq 
with the currents $J=\frac{ik}{2\pi}g\partial_-g^{-1}$ 
generating a level $N_f$ affine $SU(N_c)$ Kac-Moody algebra.
The associated energy-stress tensor 
$T^{\mu\nu}$ yields the momentum operators
\begin{eqnarray}
P^+&=&T^{++}=\frac{\pi}{N_c+N_f}\int^{\infty}_{-\infty} 
dx^- :J_{ij}(x^-)J_{ji}(x^-):\\
P^-&=&T^{+-}
=-\frac{g^2}{2}\int^{\infty}_{-\infty} 
dx^-:J_{ij}(x^-)\frac{1}{\partial^2_-}J_{ji}(x^-):.
\end{eqnarray}
To obtain the mass eigenvalues $M_n$ we have 
to solve the eigenvalue problem 
\beq\label{EVP}
2P^+P^-|\psi\rangle=M^2_n|\psi\rangle,
\eeq
which is equivalent to diagonalizing the operator $P^-$,
since $P^+$ is already diagonal. 
To discretize the system, we impose periodic boundary conditions 
of length $2L$ on the currents,
$J_{ij}(x^--L)=J_{ij}(x^-+L)$, and expand them into a discrete series of modes
\beq
J_{ij}(x^-)=\frac{1}{\sqrt{2}L}
\sum_{n=-K}^K J_{ij}(n)e^{-i\frac{2\pi n}{L}x^-},
\eeq
with
\beq\label{vacuum}
J_{ij}(n)|0\rangle=0 \quad\quad \forall n>0.
\eeq
The cutoff $K\equiv P^+L/2\pi$ 
controls the coarseness of the momentum-space discretization. 
The continuum limit is obtained
by sending $K$ to infinity. The modes of the currents obey the algebra
\begin{equation}\label{KMalgebra}
\left[J_{ij}(n),J_{i'j'}(n')\right]=n N_f \left(\delta_{ij'}\delta_{i'j}
-\frac{1}{N_c}\delta_{ij}\delta_{i'j'}\right)\delta^n_{-n'}
+\delta_{ij'} J_{i'j}(n+n')-\delta_{i'j} J_{ij'}(n+n').
\end{equation}
The momentum generators 
\begin{eqnarray}
P^+&=&\left(\frac{2\pi}{L}\right)
\frac{1}{N_c+N_f}\sum^{K}_{n=1} J_{ij}(-n)J_{ji}(n),\\
P^-&=&
\frac{\tilde{g}^2}{2\pi}\sum^{K}_{n=1} \frac{1}{n^2}J_{ij}(-n)J_{ji}(n),
\end{eqnarray}
are finite-dimensional matrices on the 
Hilbert space constructed by acting with the current operators
on the vacuum defined in Eq.~(\ref{vacuum}). For convenience we introduced 
the scaled coupling $\tilde{g}^2=\frac{g^2L}{2\pi}$. Note that the box length
$L$ drops out of the eigenvalue problem, Eq.~(\ref{EVP}).
At large $N_c$ we expect the Fock basis to consist of single-trace states
\[
\frac{1}{N_c^s}Tr[J(-n_1)J(-n_2)\cdots J(-n_s)]|0\rangle.
\] 
We note that a cyclic permutation of the currents 
will reproduce these states only up to 
states with a lower number of currents. The number of currents is not 
conserved.

\section{Analytic considerations}
\label{SecAnalytic}

The discretization of momenta via the DLCQ procedure puts severe
constraints on the possible Fock states. 
It turns out that they allow for the {\em a priori} 
calculation of two of the eigenvalues.
At harmonic resolution $K$ the states 
\begin{eqnarray}
|K\rangle&=&Tr \left[\{J(-1)\}^{K}\right]|0\rangle\\
|K-1\rangle&=&Tr \left[\{J(-1)\}^{K-2} J(-2)\right]|0\rangle
\end{eqnarray}
are unique and have $K$ and $K-1$ currents, respectively.
The sectors with less than $K-1$ currents contain more than one state, 
{\em e.g.}  
\begin{eqnarray}
|K-2\rangle_1&=&Tr \left[\{J(-1)\}^{K-3} J(-3)\right]|0\rangle,\\
|K-2\rangle_s&=&Tr \left[\{J(-1)\}^{K-2-s} J(-2)\{J(-1)\}^{s-2} 
J(-2)\right]|0\rangle, \quad 2\leq s \leq \frac{K}{2}.
\end{eqnarray}
It is relatively straightforward to derive the expressions
\begin{eqnarray}
P^-|K\rangle&=&\frac{\tilde{g}^2}{2\pi}(N_c+N_f)K |K\rangle+{\cal O}(|K-1\rangle),\\
P^-|K-1\rangle&=&\frac{\tilde{g}^2}{2\pi}(N_c+N_f)
\left(K-\frac{3}{2}\right)|K-1\rangle+{\cal O}(|K-2\rangle),\\
P^-|K-2\rangle_i&=&0+{\cal O}(|K-2\rangle),
\end{eqnarray}
where ${\cal O}(|p\rangle)$ are terms involving states 
with $p$ or less currents.
Let the dimension of the discrete Fock space be $d$ and define 
$\mu_1:=\frac{\tilde{g}^2}{2\pi}(N_c+N_f)K$ and $\mu_2:=\frac{\tilde{g}^2}{2\pi}(N_c+N_f)\left(K-\frac{3}{2}\right)$.
Then the structure of the Hamiltonian matrix is
\beq
P^-=\left(
\begin{array}{c|c}
A & 0\\\hline
B & 
\begin{array}{cc}
\mu_2 &0\\
C & \mu_1
\end{array}
\end{array}
\right).
\eeq
The matrices $A$ and $B$ have dimensions $(d-2)\times(d-2)$ and 
$2\times(d-2)$,
respectively, and $C$ is a real number. 
Clearly, $\mu_1$ and $\mu_2$ are two eigenvalues of $P^-$. 
The eigenvalues of the mass squared operator are then
\begin{equation}\label{analyticEVs}
M^2_1=\frac{g^2 N_c}{\pi}(1+\lambda)K^2  \quad\mbox{ and }\quad
M^2_2=\frac{g^2 N_c}{\pi}(1+\lambda)K\left(K-\frac{3}{2}\right).
\end{equation}
These eigenvalues seem to diverge in the continuum limit, 
which would render them physically irrelevant.
However, one can show that the eigenvalues $M^2_i(K)$ lie in different 
$Z_2$ sectors for even and odd $K$ and therefore cannot be connected.
They rather mark the appearance of new states in the spectrum, as we will see. 

%----------------------------------------------------------------------%
%           N U M E R I C A L   C A L C U L A T I O N S
%----------------------------------------------------------------------%

\section{Numerical Calculations}
\label{SecNumCalc}

%----------------------------------------------------------------------%
%                       The Hamiltonian
%----------------------------------------------------------------------%

\subsection{The Hamiltonian}
\label{SecHamiltonian}

In this section we address to calculate the eigenvalues of QCD$_2$
by solving the eigenvalue problem, Eq.~(\ref{EVP}), numerically.
Since the operator $P^+$ is already diagonal, 
we have to construct the action of the Hamiltonian $P^-$
on a basis state\footnote{A continuum version of the first step of this 
DLCQ calculation can be found in Ref.~\cite{ArmoniSonnenschein95}.}. 
Using the large $N_c$ limit implies that the basis 
states be of the 
form $Tr[J_1\cdots J_b]|0\rangle$, {\em i.e.} single-trace states. 
The current operators are subject to the Kac-Moody algebra, 
Eq.~(\ref{KMalgebra}). Annihilation operators 
may thus be produced by {\em commuting} operators.
The main obstacle for the calculations is to find a scheme to organize the 
terms in a convenient way.
We obtain such a scheme by separating terms containing
annihilation operators from those that do not.
In the definition
\begin{equation}\label{Decomposition}
\left[A,B\right]=:\lceil A,B\rceil+\lfloor A,B\rfloor,
\end{equation}
$\lceil A,B \rceil$ denotes the part of a commutator
which consists solely of creation operators.
Its complement $\lfloor A,B\rfloor$ contains annihilation operators.
It is possible to write down an expression for $P^-$ which involves only
commutators of the first type and creation operators. It reads
\begin{eqnarray}\label{actionShape}
P^- J_1 J_2 \cdots J_b|0\rangle&=&\sum_{p=1}^n
\sum^{b}_{
%k_i=1 \atop 
k_1<k_2<\ldots<k_p}
J_1\cdots J_{k_1-1}J_{k_1+1}\cdots J_{k_2-1}J_{k_2+1}\cdots J_{k_p-1} 
\nonumber\\
&& \times\lceil\lfloor\cdots \lfloor P^-,J_{k_1}\rfloor, J_{k_2}\rfloor, 
\ldots, J_{k_p} \rceil J_{k_p+1}\cdots J_b|0\rangle.
\end{eqnarray}
The commutators in this expression, and thus the 
Hamiltonian matrix, are linear in $N_f$ by
construction. 
To construct the Hamiltonian matrix, we have to 
evaluate all $2^b-1$ commutators.
In the worst case scenario  
we would have an exponentially growing number of terms 
($\frac{3}{2}\sum_{p=1}^p 2^{2(p-1)}\frac{p!b!}{(b-p)!}$) in the Hamiltonian,
due to the $\frac{3}{2}2^{2(p-1)}$
terms in a commutator involving $p$ currents.
Fortunately, the number of terms of leading power in $N_c$ grows only 
quadratically, like $(2b-1)b$.
We developed a computer code to evaluate Eq.~(\ref{actionShape})
symbolically. This task exceeds typical workstation capabilities at 
$b\ge 7$. It is however possible to deduce the expression for arbitrary
$b$, because a repeated pattern evolves.
In the large $N_c$ limit the action of $P^-$ on a state 
$|{\bf b}; n_1,\ldots,n_b\rangle:=
\frac{1}{N_c^b}J^{j_{1}}_{j_{2}}(n_1)\cdots J^{j_{b}}_{j_{1}}(n_b)|0\rangle$
with $b$ currents is then
\begin{eqnarray}\label{Pminus}
&&\!\!\!\!\!\!\!\!\!\!\!\!\!\!P^-|{\bf b}; n_1,\ldots,n_b\rangle\\
% term #1
&=&-\frac{\tilde{g}^2N_c}{2\pi}\sum_{i=1}^b\left(\sum_{m=1}^{n_i-1}\frac{1}{(m-n_i)^2}-
\sum_{m=1}^{n_i-1}\frac{1}{m^2}\right)
|{\bf b+1};n_1,n_2,\ldots,n_i-m,m,\ldots, n_b\rangle\nonumber\\
% term #2
&+&\frac{\tilde{g}^2N_c}{2\pi}\sum_{i=1}^b\left(\frac{\lambda}{n_i}
+\sum_{m=1}^{n_i-1}\frac{1}{m^2}\right)
|{\bf b};n_1,n_2,\ldots, n_b\rangle\nonumber\\
% term #3
&+&\frac{\tilde{g}^2 N_c}{2\pi}\sum_{i=1}^{b-1}\left(\sum_{m=0}^{n_{i+1}-1}
\frac{1}{(m+n_i)^2}-\sum_{m=1}^{n_{i+1}-1}\frac{1}{m^2}\right)
|{\bf b};n_1,n_2,\ldots,n_i+m,n_{i+1}-m,\ldots, n_b\rangle\nonumber\\
&+&\frac{\tilde{g}^2 N_c}{2\pi}\left(\sum_{m=0}^{n_b-1}\frac{1}{(m+n_1)^2}-
\sum_{m=1}^{n_b-1}\frac{1}{m^2}\right)
|{\bf b};n_2,\ldots,n_{b-1},n_b-m,n_{1}+m\rangle\nonumber\\
%
% term #4
%
&+&\lambda\frac{\tilde{g}^2 N_c}{2\pi}\sum_{j=1}^{b-2}\left\{\right.
%
%  term 4.1
(-)^j\sum_{i=1}^{b-j}\left[
\frac{1}{(\sum_{q=i}^{j+i}n_q)^{2}}-
\frac{1}{(\sum_{q=i+1}^{j+i}n_q)^{2}}
\right]n_{i+j}\nonumber\\
&&\quad\quad\quad\quad\times
|{\bf b-j};n_1,\ldots,n_{i-1},\sum_{q=i}^{j+i}n_q,n_{j+i+1},
\ldots, n_b\rangle\nonumber\\
%
%  term 4.2
&&\quad\quad\quad(-)^{j-1}\left[
\frac{1}{(n_1+\sum_{q=b-j+1}^{b}n_q)^{2}}-
\frac{1}{(\sum_{q=b-j+1}^{b}n_q)^{2}}
\right]n_{b}\nonumber\\
&&\quad\quad\quad\quad\times
|{\bf b-j};n_2,\ldots,n_{b-j},n_1+\sum_{q=b-j+1}^{b}n_q\rangle\nonumber\\
%
%  term 4.3
&&\quad\quad\quad(-)^j\sum_{i=1}^{j-1}\left[
\frac{1}{(n_1+\sum_{q=b-i+1}^{b}n_q)^{2}}-
\frac{1}{(\sum_{q=b-i+1}^{b}n_q)^{2}}
\right]n_{b}\nonumber\\
&&\quad\quad\quad\quad\times
\left[
|{\bf b-j};n_{j-i+1},n_{j-i+2},\ldots,n_{b-i-1},\sum_{q=1}^{j-i}n_q
+\sum_{q=b-i}^{b}n_q\rangle\right.\nonumber\\
&&\quad\quad\quad\quad\quad\left.
-\left.|{\bf b-j};n_{j-i+2},n_{j-i+3},\ldots,n_{b-i},\sum_{q=1}^{j-i+1}n_q
+\sum_{q=b-i+1}^{b}n_q\rangle
\right]\right\}\nonumber\\
%
% term #5
&&+\frac{\tilde{g}^2N_c}{2\pi}\sum_{j=1}^{b-2}\left\{
%
% term 5.1
%\quad\quad\quad
(-)^j\sum_{i=1}^{b-j-1}\sum_{m=0}^{n_{i+j+1}-1}\left(
\frac{1}{(m+\sum_{q=i}^{i+j}n_q)^2}-
\frac{1}{(m+\sum_{q=i+1}^{i+j}n_q)^2}\right)\right.\nonumber\\
&&\quad\quad\quad\quad\quad\times
|{\bf b-j};n_1,n_2,\ldots,\sum^{i+j}_{q=i}n_{q}+m,n_{i+j+1}-m,
n_{i+j+2},\ldots, n_b\rangle\nonumber\\
%
% term 5.2
&&\quad\quad(-)^j\left(\sum_{m=0}^{n_{b}-1}
\frac{1}{(m+n_1)^2}-
\sum_{m=1}^{n_{b}-1}\frac{1}{m^2}\right)
%\nonumber\\
%&&\quad\quad\quad\quad\times
|{\bf b-j};n_{j+2},\ldots,n_{b-1},n_b-m,\sum_{q=1}^{j+1}n_{q}+m\rangle
\nonumber\\
%
% term 5.3
&&\quad\quad(-)^{j-1}\left(\sum_{m=0}^{n_{b}-1}
\frac{1}{(m+n_1)^2}-
\sum_{m=1}^{n_{b}-1}\frac{1}{m^2}\right)
%\nonumber\\
%&&\quad\quad\quad\quad\times
\left.
|{\bf b-j};n_{j+1},\ldots,n_{b-1},\sum_{q=1}^j n_q+n_b\rangle\right\}
\nonumber
\end{eqnarray}
Note that only the terms in lines two and five of this result
contain $N_f$. These terms will be
absent in the 't Hooft limit ($N_f/N_c\rightarrow 0$) and will be dominant
in the large $N_f$ limit.

\begin{table}
\centerline{
\begin{tabular}{|c|ccccccccccccc|}\hline
$K$ & 2 & 3 & 4 & 5 &6 & 7& 8 & 9 & 10 & 11 & 12 & 13 &14\\\hline
states & 1 &2 & 4 & 6 &12& 18 &34 &58 &106 & 186 & 350 & 630 & 1180\\
\hline
\end{tabular}
}
\caption{Number of basis states a a function of the harmonic resolution $K$.}
\label{TableStates}
\end{table}
%\vspace{0.5cm}

\subsection{Numerical results}

We are solving the eigenvalue problem, Eq.~(\ref{EVP}), numerically 
to obtain the mass 
spectrum. Recall that we have two parameters at our disposal to study 
the spectra. One is the harmonic resolution $K$, which we are supposed to 
send to infinity. The other is the ratio of the numbers of flavors and 
colors, $\lambda=N_f/N_c$, of the fermions in the theory. 
The most prominent cases are $\lambda=0,1,\infty$, namely,
the 't Hooft limit, adjoint fermions, and the large $N_f$ limit.
Note that also the unphysical parameter $K$ might give insight into the 
spectrum, {\em e.g.}~the discovery of continuum states via their 
characteristic $K$ dependence in Ref.~\cite{GHK97}.

The advantages of formulating the problem with
$SU(N)$ currents rather than with fermionic degrees of freedom are 
twofold. For one, the much smaller basis allows for a larger resolution
$K$ and thus for more accurate results. We will also 
see the structure of the spectrum much clearer, in the sense that 
many, if not all, of the uninteresting multi-particle states are absent. 
Secondly, we are able to
study the behavior of the spectrum as we couple the gauge field to different
forms of matter by varying the parameter $N_f/N_c$, which is an algebraic 
variable in the present approach. 
This might be used a tool to interpret spectra.
% when 
%trying to understand mechanisms such as screening.
% and transitions from 
% single to multi-particle states.

Performing the numerical calculation, we obtain exactly the same 
eigenvalues in the adjoint case as in previous 
works \cite{BhanotDemeterfiKlebanov93} with {\em anti-periodic} boundary 
conditions for the 
fermions. The number of states grows exponentially with the harmonic 
resolution, {\em cf.}~Table \ref{TableStates}.
At $K=12$ we are diagonalizing a Hamiltonian of 
dimensions\footnote{It is not clear {\em a priori} how to construct 
a Fock basis for the current operators. 
A naive expectation is that single-trace states modulo
cyclic permutations provide such a basis.
We convinced ourselves that this is actually the correct choice 
by explicitly performing a large $N_c$ 
Gram-Schmidt orthonormalization for small $K$ on these states.}
$350\times 350$, 
whereas in the fermionic approach one would have to operate
on a Fock space with 4338 states to obtain the same accuracy.
To further reduce the computational effort, we can use the 
$Z_2$ symmetry of the Hamiltonian which is invariant under the 
transformation
${\cal{T}}J_{ij}(n)=-J_{ij}(n)$.   
%%%%%%%%%%%%%%%%%%%%%%%%%%%%%%%%%%%%%% Changed 10/7/00 %%%%%%%%%%%%%%
It is straightforward to convince one-self that the action of this
operator on a state with $b$ currents is
\begin{equation}
{\cal T}|{\bf b}; n_1,n_2,\ldots,n_b\rangle=
(-)^b\sum_{i=0}^{2^{b-2}}|{\bf p_i}; n_1,T_i(n_b,n_{b-1},\ldots, n_2)\rangle,
\end{equation}
where the $T_i$ consist of $p_i$ partial sums of 
the $b-1$ momenta, in the sense that
$i$ runs over all possibilities to place $0,1,\ldots, b-2$ commas
between the momenta while summing those momenta which are not
separated by a comma, {\em e.g.} 
\begin{eqnarray}
{\cal T}|{\bf 4}; n_1,n_2,n_3,n_4\rangle&=&|{\bf 4}; n_1,n_4,n_3,n_2\rangle
+|{\bf 3}; n_1,n_4+n_3,n_2\rangle\nonumber \\
&&+|{\bf 3}; n_1,n_4,n_3+n_2\rangle+|{\bf 2}; n_1,n_4+n_3+n_2\rangle.
\end{eqnarray}
In order not to complicate the construction of the Hamiltonian, we
determine the $Z_2$ parity of an eigenstate {\em a posteriori} 
by calculating the expectation value of the operator ${\cal T}$ 
in this state.
%%%%%%%%%%%%%%%%%%%%%%%%%%%%%%%%%%%%%%%%%%%%%%%%%%%%%%%%%%%%%%%%%%%%%
The separation of the $Z_2$ odd and even eigenfunctions is useful when
interpreting the results, because it reduces the density of eigenvalues to
roughly a half.

\begin{figure}
\centerline{\psfig{figure=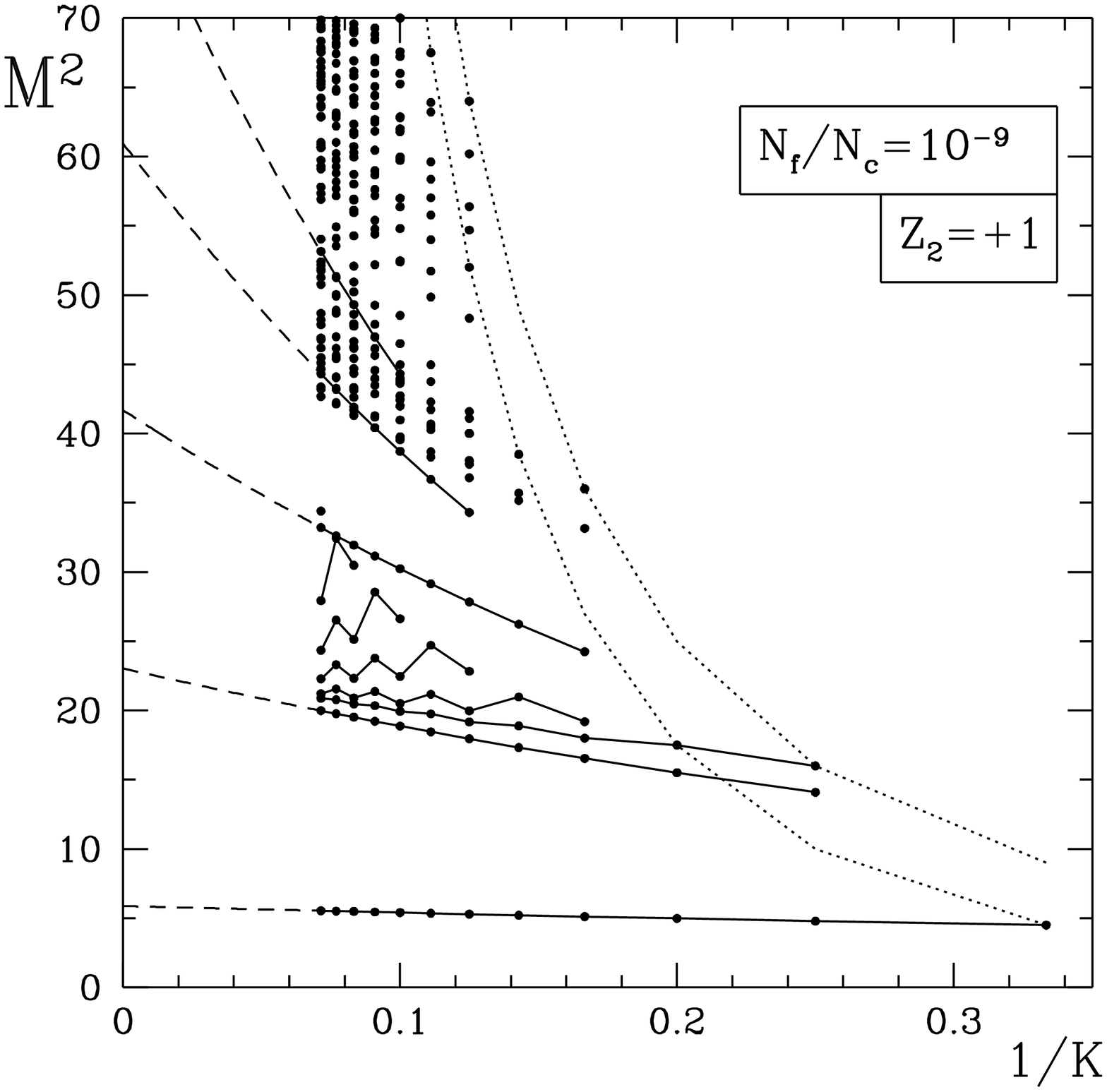,width=8cm,angle=0}
\psfig{figure=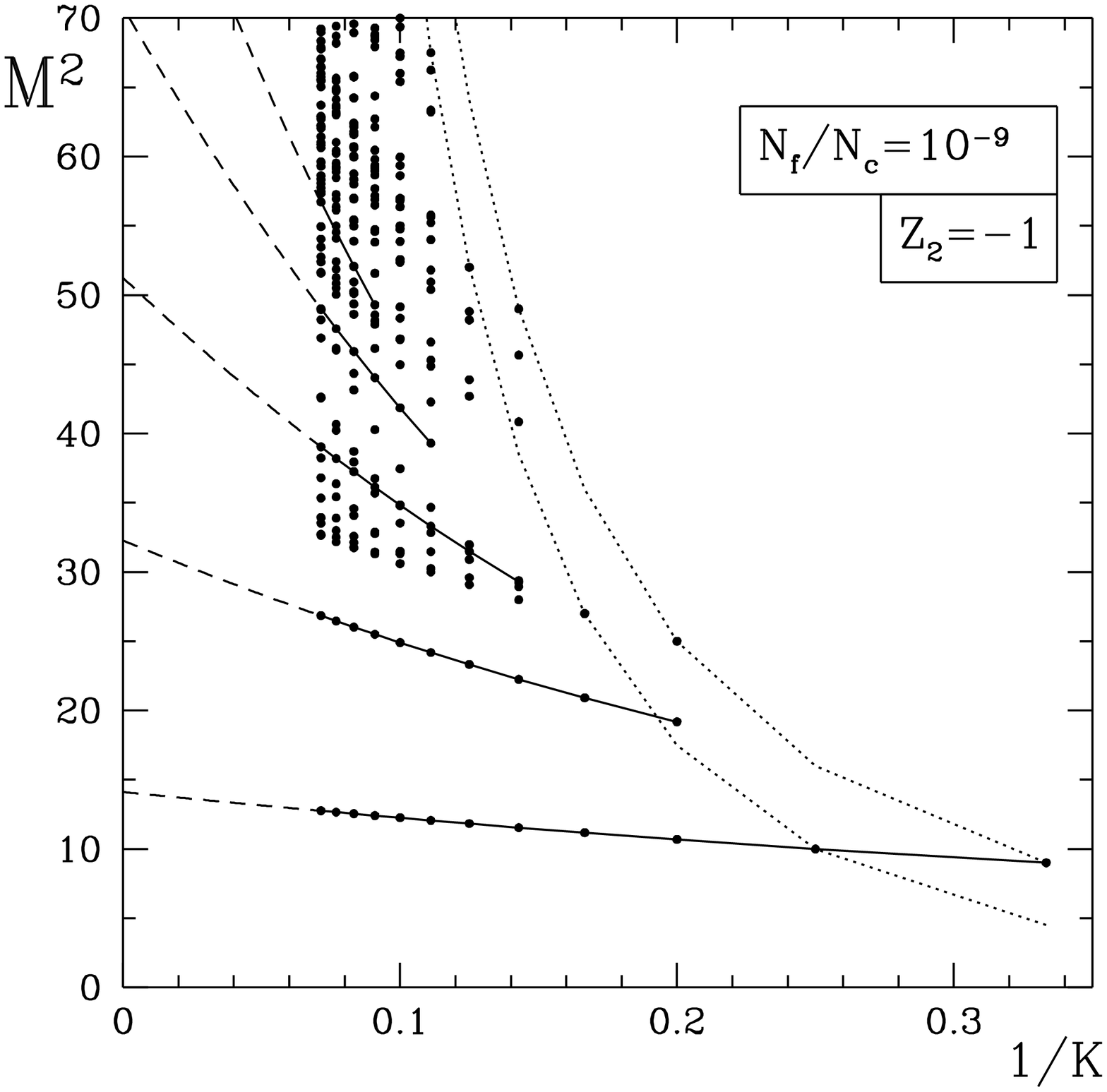,width=8cm,angle=0}
}
\caption{Spectra of the 't Hooft limit in the $Z_2$ even (a) and
odd (b) sectors. Solid lines connect associated eigenvalues as $K$ varies.
Short dashed lines connect analytically calculable eigenvalues. Long dashed 
lines  are extrapolations towards the continuum limit. Masses are in units 
$g^2N_c/\pi$.}
\label{tHooftSpectrum}
\end{figure}
\noindent

\subsection{The 't Hooft limit}

To test the consistency of our approach, we compare to the well-known
results in the large $N_c$ and the large $N_f$ limit. 
We find complete agreement. First let us consider the large $N_c$
(or 't Hooft) limit,
where we should recover the results of the 't Hooft model \cite{tHooft74b}.
The spectrum of large $N_c$ QCD in two
dimensions with massless fundamental fermions asymptotically has the form 
$
M^2=\pi^2 n
$
for large integer $n$, where the mass squared is 
in units ${g^2N_c}/{\pi}$. We find that at $K=14$ the lowest ten
single-particle states have 
masses\footnote{We performed the continuum limit here and in 
the sequel by fitting the data to a polynomial of second degree in $1/K$.} 
\beq
M^2= 5.88,14.11,23.04,32.27,41.68,51.24,60.93,70.76,80.97,90.90,
\eeq
which is in very good agreement with 't Hooft's numerical solution 
\cite{tHooft74b}.
%\[
%M^2_{\rm tH}= 5.87,14.09,23.03,32.25,\ldots
%\]
The actual spectrum, 
Fig.~\ref{tHooftSpectrum}, is a mixture of single-
and multi-particle states. The multi-particle states decouple completely 
from the single-particle states. 
They appear here, because we do not use an orthonormal basis. 
We have performed a
calculation with an orthonormal set of states up to $K=6$ and found 
that then only single-particle states are present.
We were able to identify all multi-particle states as composites of two or 
more single-particle 
states. The masses $M^2_{mp}$ of multi-particle states are
given by \cite{GHK97}
\begin{equation}\label{MultiParticleMass}
\frac{M^2_{mp}(K)}{K}=\frac{M^2_{p_1}(n)}{n}+\frac{M^2_{p_2}(K-n)}{K-n},
\end{equation}
where $M^2_{p_i}(n)$ are the masses of single-particle states at harmonic 
resolution $n$. These
masses are exactly reproduced in the spectrum. 
This identification in turn made it possible to detect the single-particle 
states hidden amongst the multi-particle states.
%%%%%%%%%%%%%%%%%%%% Change 13/7/00 %%%%%%%%%%%%%%%%
Surprisingly, the eigenfunctions have no significant structure,
apart from the special form of the lowest state in each $Z_2$
sector. In particular, we were unable to distinguish  
single- from multi-particle wavefunctions by their shapes.
%%%%%%%%%%%%%%%%%%%%%%%%%%%%%%%%%%%%%%%%%%%%%%%%%%%%
%An interesting observation is that the multi-particle state made out of 
%the lightest boson $B_1$ and the third-lightest $B_3$ lie in different 
%$Z_2$ sectors of the theory. Setting $p_1=B_1$ and $p_2=B_3$, we get 
%masses of the $Z_2$ even sector, whereas exchanging the particles gives
%masses in the odd sector. This could be a hint to the coupling of the massless
%sector of the theory. 
%Recall that Gross et al \cite{GHK97} and especially Antonuccio and Pinsky 
%\cite{AntonuccioPinsky98}
%find this behavior (exactly matching multi-particle states) in the adjoint 
%case. Additionally they find associated with every exactly matching doublet 
%another state a bit off the expected mass \cite{AntonuccioPinsky98}.  
%We will see in the next section that we have only the latter states 
%in our spectrum. This means
%that these states are not decoupled in the adjoint case, 
%whereas the states exactly at the expected masses are true
%multi-particle states. Here, we conclude likewise that the exactly predicted 
%states are indeed multi-particle states. 
%The question remains, why the multi-particle states appear in a 
%single-trace basis. A possible answer, inspired by Antonuccio and Pinsky's
%finite $N$ analysis \cite{AntonuccioPinsky98}, could be that the single-trace 
%multi-particle
%states come together with a multi-trace state of the same mass, which we 
%don't see in our large $N$ calculation. According to this interpretation, 
%we see here an artifact of finite $N$. 

\begin{figure}
\centerline{
\psfig{figure=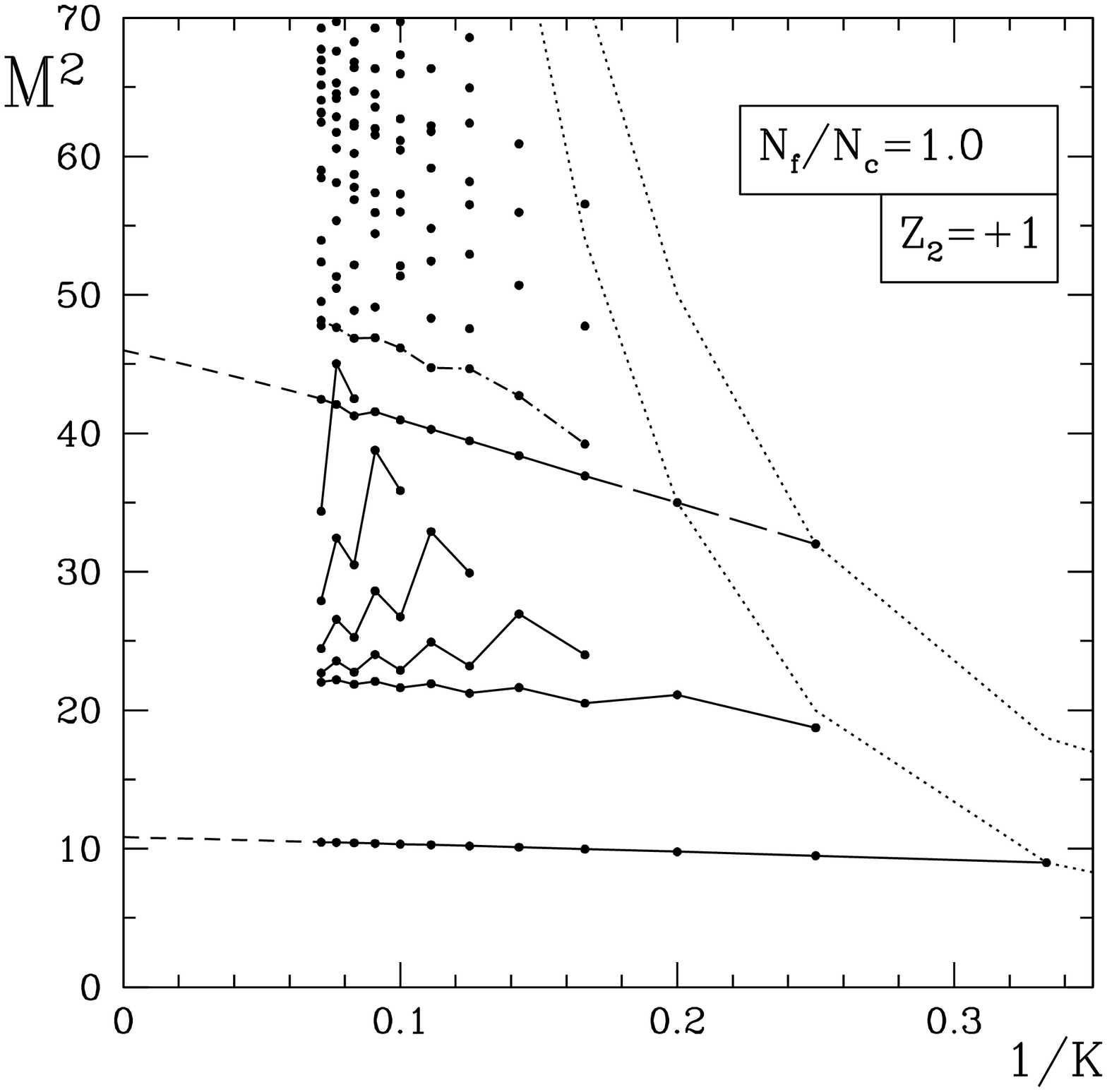,width=8cm,angle=0}
\psfig{figure=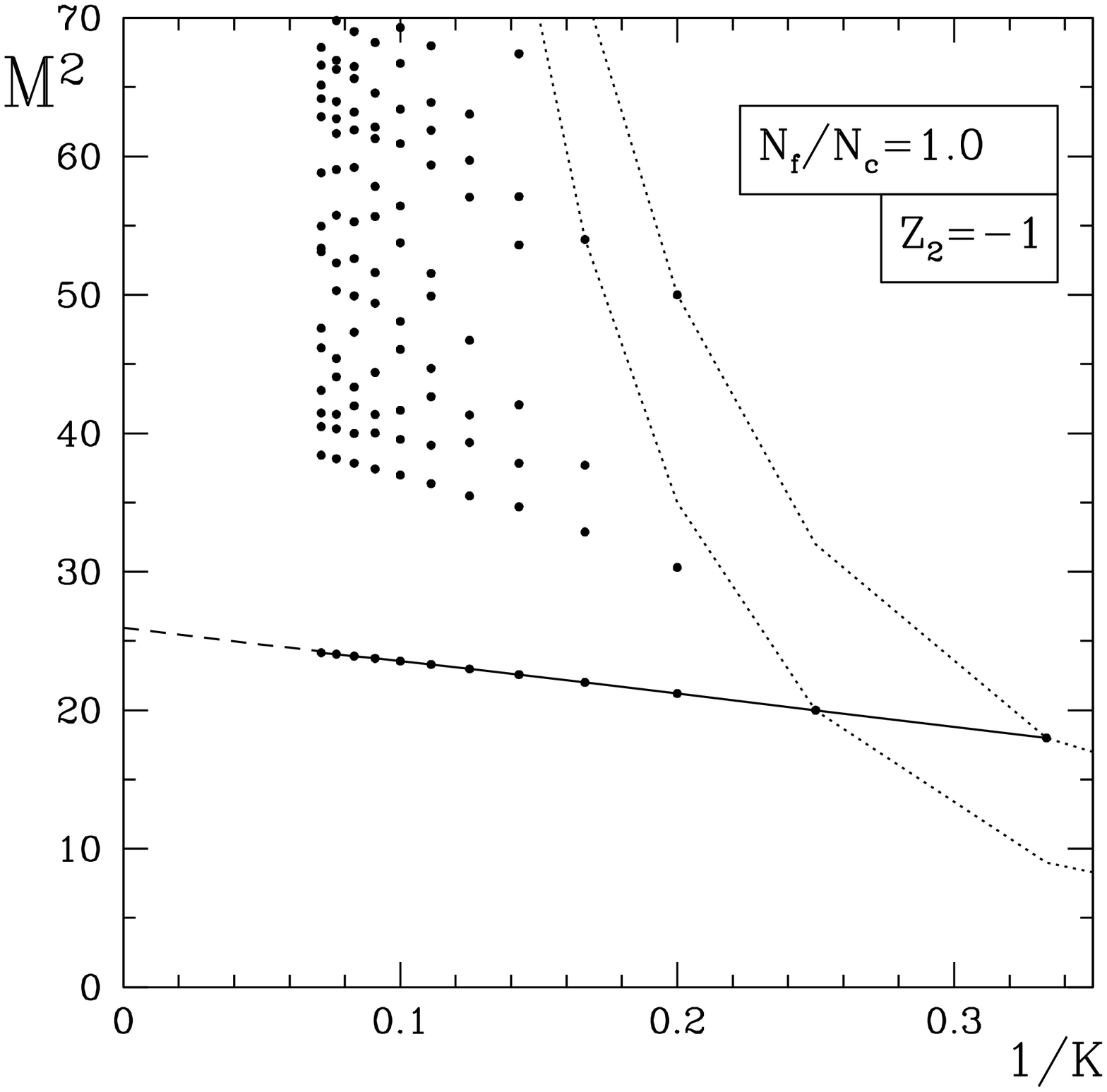,width=8cm,angle=0}
}
\caption{Spectra of the theory with adjoint fermions in the $Z_2$ even (a) and
odd (b) sectors. Solid lines connect associated eigenvalues as $K$ 
varies. Short dashed lines connect analytically calculable eigenvalues.
Dashed lines  are extrapolations towards the continuum limit.
Masses are in units $g^2N_c/\pi$. The dash-dotted trajectory is the 
analog of the third single-particle state in the 't Hooft limit.}
\label{AdjointSpectrum}
\end{figure}
%%%%%%%%%%%%%%%%%%%%%%%%%%%%%%%%%%%%%5

\subsection{Adjoint fermions}

Consider now the adjoint spectrum, Fig.~\ref{AdjointSpectrum}.
If we look at the eigenvalue trajectories (mass squared as a function of $K$),
the structure of the spectrum looks similar to the \mbox{'t Hooft} case.
We see immediately three single-particle candidates 
which qualify by their straight, smooth trajectories. 
In the continuum limit they have the eigenvalues
\beq
M^2_{B_1}=10.84, M^2_{B_2}=25.73, M^2_{B_3}=45.66.
\eeq
The expectation for the asymptotic solution is here \cite{Kutasov94}
\beq
{M}^2_{n_1,n_2,\ldots,n_k}= 2\pi^2(n_1+n_2+\cdots+n_k);\quad
n\in 2\mbox{\bf Z},
%=4\pi^2, 8\pi^2, 12 \pi^2,\ldots,
\eeq
again in units ${g^2N_c}/{\pi}$.
The single-particle masses appear to be at roughly 
twice the values of the 't Hooft limit, whereas the low-lying 
'multi-particle'\footnote{We use quotation marks here and in the 
following because it is not {\em a priori} clear that these states are 
indeed multi-particle states.} masses 
stay at more or less their 't Hooft values.
The crucial difference between the spectra is that the 
'multi-particle' eigenvalues of the adjoint case coincide only approximately
with the values from the mass formula, Eq.~(\ref{MultiParticleMass}). 
They are systematically lower than expected. 
As such, these states appear to be loosely 
{\em bound} composites, {\em i.e.} they 
have to be interpreted as single-particle states, at least at finite $K$. 
We can see that these states are a vital part of the spectrum by
comparing the eigenvalues at $M^2\simeq 32, K=13$ in 
Fig.~\ref{tHooftSpectrum}(a) with those of Fig.~\ref{AdjointSpectrum}(a) at
$M^2\simeq 42, K=12$. In the latter case the eigenvalues repel each other,
leading to a deformation of the eigenvalue trajectory of the (conjectured)
single-particle state $B_3$. This means that 
the 'multi-particle' states
are interacting! 
By varying the parameter $\lambda$, we were able to 
follow this deformation. As $N_f$ grows, the lowest  
multi-particle states of the 't Hooft spectrum 'move through' 
the single-particle 
state with mass squared $M^2=23.04$ in the 't Hooft limit,
and produce these deformations in a trajectory.
In the 't Hooft limit, 
Fig.~\ref{tHooftSpectrum}(a), we see no distortion of the single-particle 
trajectory though the corresponding eigenvalue is almost degenerate with
a (decoupled) multi-particle state\footnote{The same 
deformation occurs in the fermionic spectrum, 
namely in the trajectory of the second $Z_2$ even single-particle 
state of Ref.~\cite{GHK97}.}.

The question arises, whether these loosely bound states become 
multi-particle states and decouple 
in the continuum limit in the adjoint spectrum. For the states between 
$M^2=20$ and $M^2=40$ in the $Z_2$ even sector, the answer is certainly yes.
We performed a fit to the data finding 
that the eigenvalues converge towards a single point, namely 
the two particle continuum threshold at $M^2=22.85=4 M^2_{F_1}$, {\em i.e.}
twice the mass of the lowest fermionic state \cite{GHK97}.
The deviations from the mass formula, Eq.~(\ref{MultiParticleMass}), 
vanishes faster than 
$\frac{1}{K^{\beta}}$, where $\beta>2$.

%%%%%%%%%%%%%%%%%%%%%%%%%%%%%%%%%%%%%%%%%%%%%%%%%%%%%%%%%%%%%%
\begin{figure}
\centerline{
%%%%%%%%%%%%%%%%%%%%%%%%%%%%%
% Left EFs (8 X 2001)
%%%%%%%%%%%
%\psfig{figure=multiSF.ps,width=8cm,angle=0}
%\psfig{figure=multiWF.ps,width=8cm,angle=0}
%%%%%%%%%%%%%%%%%%%%%%%%%%%%%
% Right EFs (8 X 2001)
%%%%%%%%%%%
\psfig{file=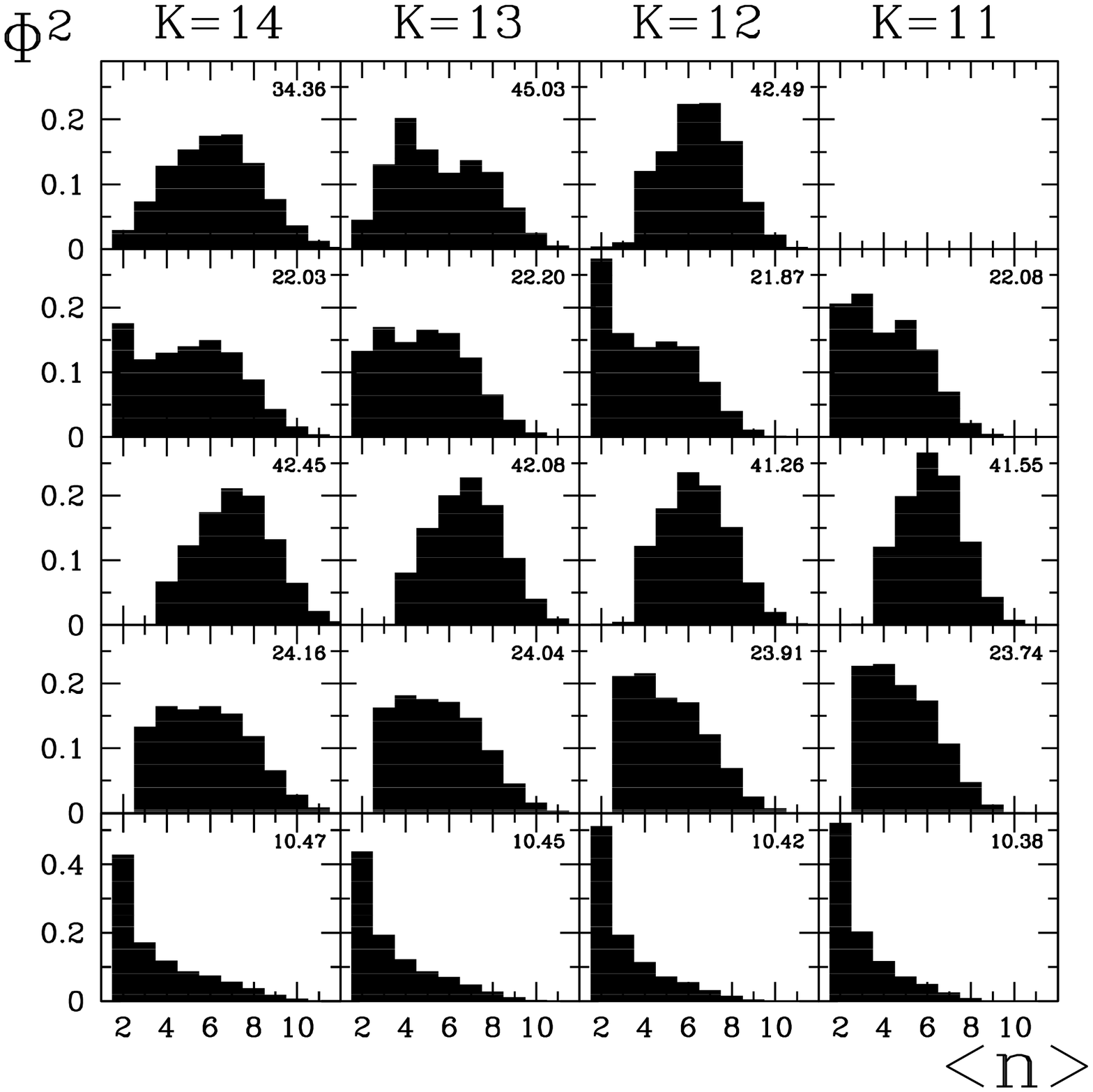,width=8cm}
\psfig{file=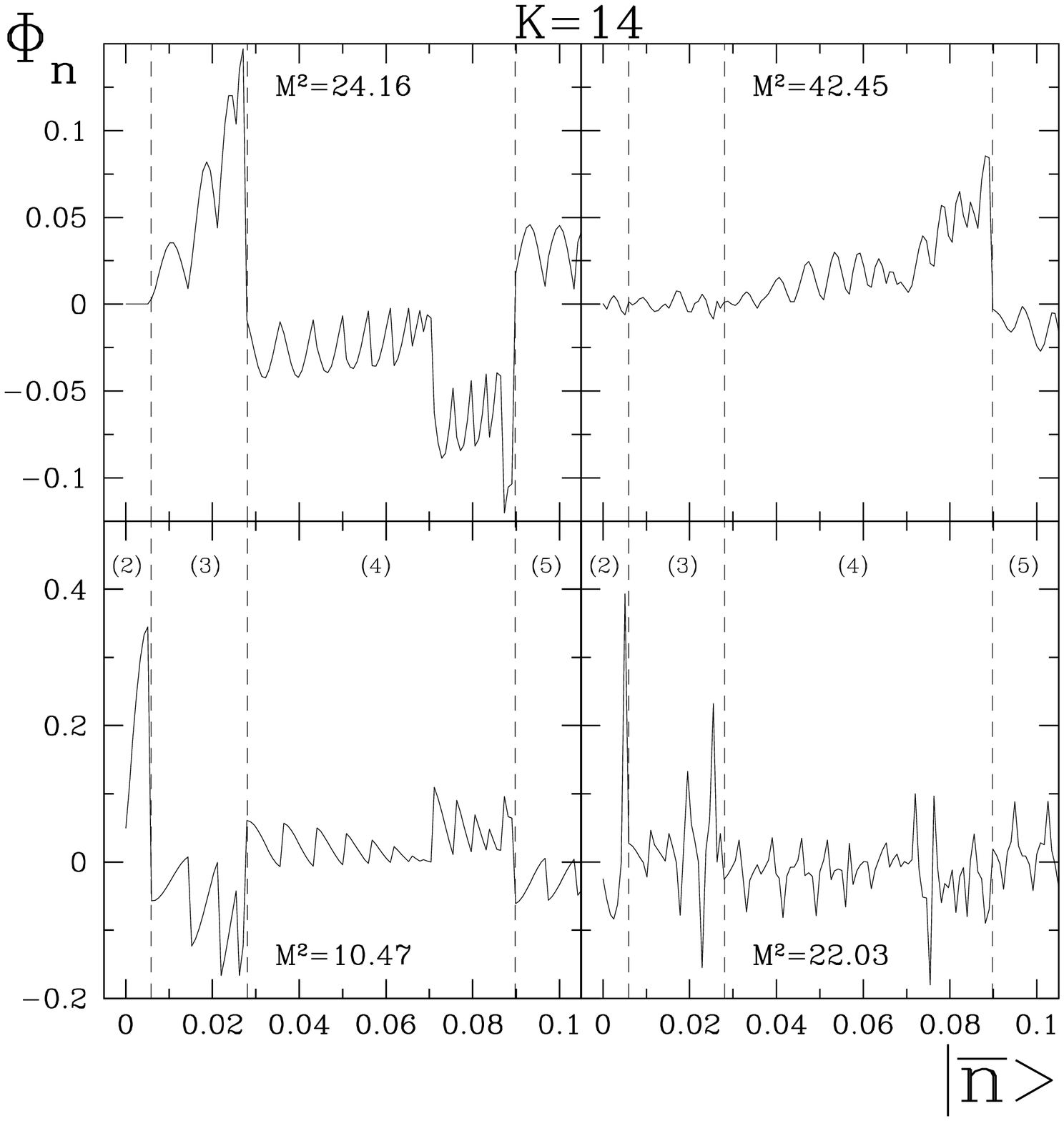,width=8cm}
}
\vspace*{-0.5cm}
\caption{Adjoint fermions: 
Left (a): The current number distribution functions 
of selected states at different 
harmonic resolution $K$. The lower three rows are
distributions of (conjectured) single-particle states. The state in the
top row is not present at $K=11$. Right (b): Wavefunctions of selected
eigenstates at $K=14$ as functions of the basis state number divided by the 
total number of states (1180). The dashed lines mark points
of changing parton content (number of currents in parenthesis).}
\label{AdjointSFs}
\end{figure}
%%%%%%%%%%%%%%%%%%%%%%%%%%%%%%%%%%%%%%%%%%%%%%%%%%%%%%%%%%%%%%%

The deformation of a trajectory renders it questionable if the associated
state is
really a single-particle state. As $K$ grows there will be more
and more 'multi-particle' states to distort the trajectory. On the other
hand, the 'multi-particle' sector will interact less and less as $K$ increases.
The number of (low-lying) 'multi-particle' states grows linearly with $K$.
If we can take the deviation of the 'multi-particle' states from the 
mass formula (proportional to $\frac{1}{K^{\beta}}$) 
as a measure of the coupling, 
we have to conclude that the single-particle trajectory will be 
less and less distorted and thus a true single-particle state will emerge
as $K\rightarrow \infty$. 
Although the first two points of the 
trajectory at $K=4$ and $K=5$ are multi-particle states with fermionic 
constituents\footnote{One state consists of $M^2_{F_1}(K{=}\frac{5}{2}){=}5$ 
and $M^2_{F_2}(K{=}\frac{5}{2}){=}12.5$, the other of the first mass plus 
$M^2_{F_2}(K{=}\frac{3}{2}){=}9$.}, we feel quite safe about the single 
particle nature of this state, because we find that 
the same situation occurs in the case of 
the lowest $Z_2$ odd trajectory, which represents undoubtedly a single-particle
state, due to its complete isolation in the spectrum.
To check whether all 't Hooft mesons become adjoint single-particle states,
we followed the next two $Z_2$ even 
%%%%%%%%%%%%%
single-particles 
%%%%%%%%%%% changed 8X 2001 from %%%%%%%%%%%
%single-particle trajectories
%%%%%%%%%%% 
of the 't Hooft spectrum
as we varied $\lambda$ in small steps. One is shown as the dashed-dotted 
line in Fig.~\ref{AdjointSpectrum}
just above the state converging to $M^2{=}45.99$. 
Their irregular shape makes 
it unlikely that they are single-particle states. 

%%%%%%%%%%%%%%%%%%%%%%%%%%%%%%%%%%%%%%%%%%%%%%%%%%%%%%%%%%%%%%
\begin{figure}
\centerline{
\psfig{figure=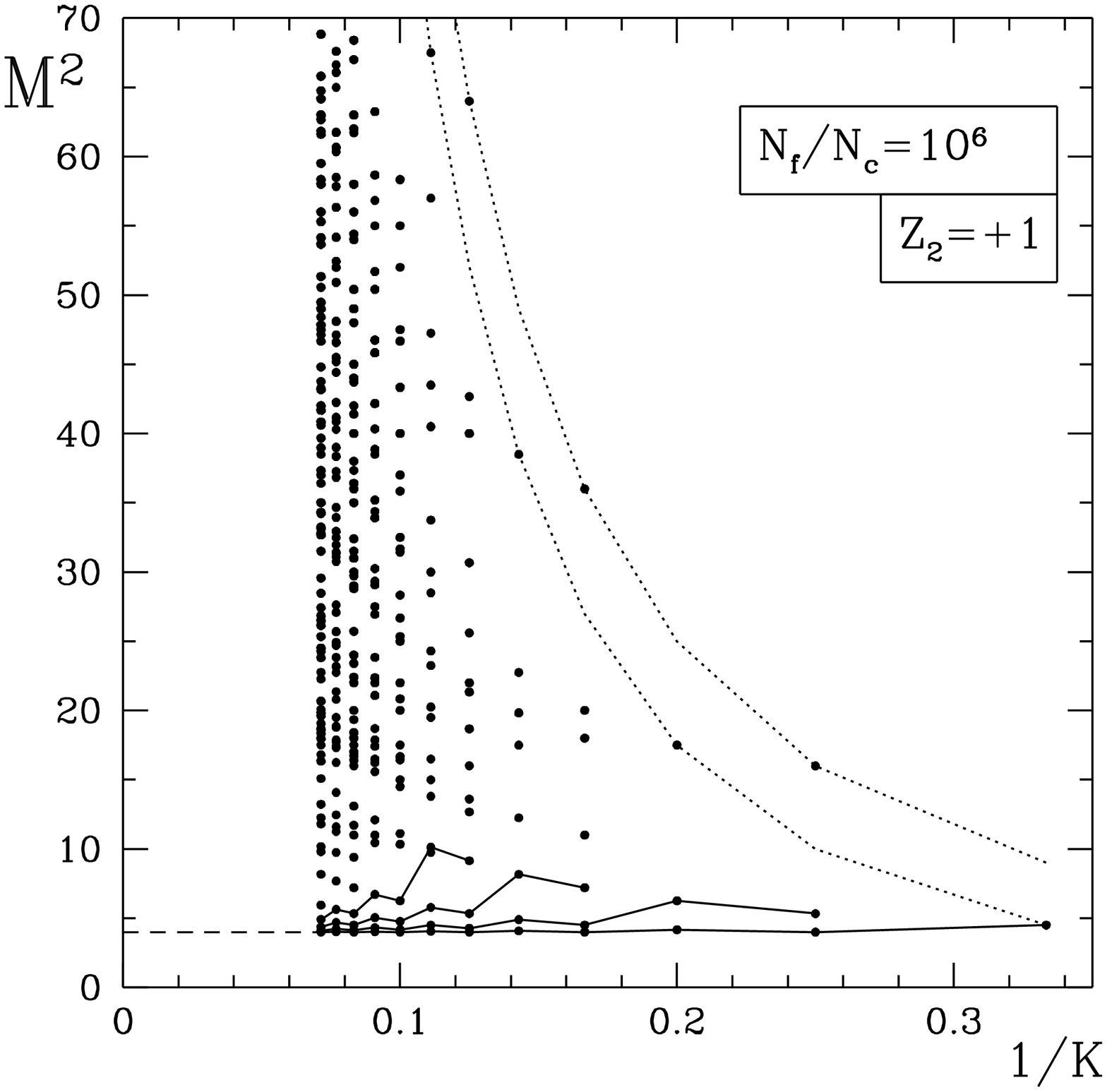,width=8cm,angle=0}
\psfig{figure=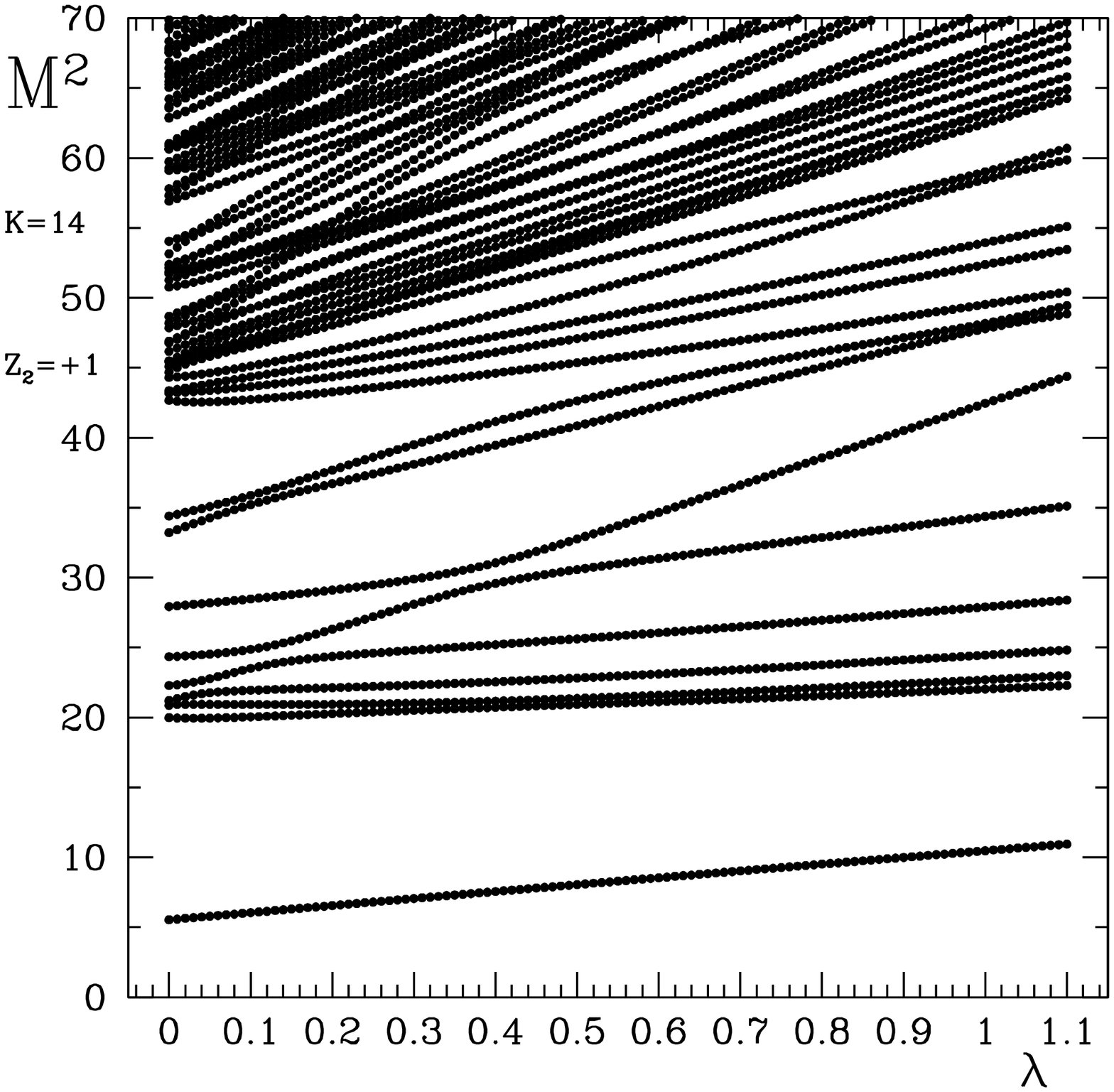,width=8cm,angle=0}
}
\vspace*{-0.45cm}
\caption{Left (a): Spectrum of the $Z_2$ even sector in large $N_f$ limit.
Note that masses are in units $g^2N_f/\pi$.
The dashed line is the extrapolation 
of the lowest eigenvalue towards the continuum. 
Right (b): Mass squared eigenvalues as
functions of $\lambda=N_f/N_c$. 
Masses are in units $g^2N_c/\pi$.
}
\label{largeNSpectrum}
\vspace*{-0.3cm}
\end{figure}
%%%%%%%%%%%%%%%%%%%%%%%%%%%%%%%%%%%%%%%%%%%%%%%%%%%%%%%%%%%%%%

The eigenfunctions and the probability distributions in parton number were
helpful to identify states at different $K$ in the fermionic formulation of
the theory \cite{DalleyKlebanov93b,GHK97}. 
Here, with a non-conserved parton number, these quantities 
will be of limited
usefulness. In the fermionic formulation
the single-particle states have a high probability to be
states of a definite parton number \cite{DalleyKlebanov93b}. 
%%%%%%%%%%%%%%%%%%%%%%%%%%%%
% Misleading below (8 X 2001)
%%%%%%%%%%%%%%% 
%In the present work, we
%find this behavior only for the lowest state in the $Z_2$ even sector,
%{\em cf.~}Fig.~\ref{AdjointSFs}(a).
%This state is a two current state to a very good approximation ($98.4\%$).
%The next-to-lowest state is an almost
%perfect mixture of two and three parton states, although this ratio 
%varies with $K$.
%%%%%%%%%%%%%%%%%%%%%%%%%%%%
% Misleading above/Correction below (8 X 2001)
%%%%%%%%%%%%%%%%%%%%%%%%%%%% 
In the present work, we
find this behavior only for the highest states in each $Z_2$ sector,
which are states with $K$ or $K-1$ currents, 
{\em cf.~}Sec.~\ref{SecAnalytic}. 
Note, however, that the conjectured lowest 
single particle states $B_1, B_2$, and $B_3$ 
have a large two-current contribution, no two-current, and no two- and 
three-current contribution, respectively.
%%%%%%%%%%%%%%%%%%%%%%%%%%%%
% Correction above (8 X 2001)
%%%%%%%%%%%%%%% 
We definitely cannot use information about the probability distributions to 
distinguish states:
if we have a close encounter of two eigenvalues in the spectrum,
one of which is presumably a single- and the other one a multi-particle 
state, we find that the probability distributions are almost identical. 
%We observe
%a deformation of the probability distribution as we vary $K$ and approach
%the point of the close encounter. 
For the higher states significant structures in the parton probability 
distributions are missing altogether.
%%%%%%%%%%%%%%%%%%%%%%
% Slight change below (8 X 2001)
%%%%%%%%%%%%%%%%%%%%%%
Concerning the wavefunctions, we find that the lowest eigenstate has 
a peak in the two-particle sector,
{\em cf.~}Fig.~\ref{AdjointSFs}(b),
%%%%%%%%%%%%%%%%%%
% Addition to clarify misleading statements (8 X 2001)
%%%%%%%%%%%%%%%%%%
and has some half-sine structure in higher current sectors.
%%%%%%%%%%%%%%%%%%
% Addition above (8 X 2001)
%%%%%%%%%%%%%%%%%%
We find 
no obvious structure in other states that could give a hint of how to
construct solutions of the theory analytically.
%%%%%%%%%%%%%%%%%%%%%%%%%%%% 
However, there might be  
hints of an additional structure in the eigenfunctions:
we found several wavefunctions which have a noticeable amplitude drop at the
boundaries of parton number sectors.
%%%%%%%%%%%%%%%%%%%%%
As an example, consider the drop of the 
wavefunctions of the 'single-particle' states 
($B_1,B_2, B_3$) at the boundaries of the 2, 3, and 4 current sectors, 
respectively, in Fig.~\ref{AdjointSFs}(b). 
%%%%%%%%%%%%%%%%%%
% Addition to clarify misleading statements (8 X 2001)
%%%%%%%%%%%%%%%%%%
In general, 'single-particle' states exhibit a sharp change of 
the sign of the amplitude at the boundaries of current sectors.
%%%%%%%%%%%%%%%%%%
% Addition above (8 X 2001)
%%%%%%%%%%%%%%%%%%
The 'multi-particle' state ($M^2=22.03$) does not exhibit such a behavior.
%%%%%%%%%%%%%%%%%%%%%%%%%%%%%%%% Changed 11/7/00 %%%%%%%%%%%%%%%%%%%%%%%%%%
In principle, the wavefunction should factorize if it is a continuum state
consisting of two non-interacting bound states. Disentangling wavefunctions 
along these lines has been attempted in Ref.~\cite{Hornbostel}; here, however,
this task seems intractable, due to the non-conserved parton number. 
%%%%%%%%%%%%%%%%%%%%%%%%%%%%%%%%%%%%%%%%%%%%%%%%%%%%%%%%%%%%%%%%%%%%%%%%%%%

\begin{table}
\centerline{
\begin{tabular}{|c||c|c||c|c||c|}\hline
\rule[-3mm]{0mm}{8mm}&\multicolumn{2}{|c||}{$N_c=N_f$}
&\multicolumn{2}{|c||}{$N_f/N_c\rightarrow 0$}
&{$N_f/N_c\rightarrow \infty$}\\
\hline
\rule[-3mm]{0mm}{8mm}
$K$ & $M^2_{B_1}$ & $M^2_{B_2}$ & $M^2_{B_1}$ & $M^2_{B_2}$ & $M^2_{cont.}$ \\
\hline\hline
3 &  9.0000 & 18.0000 & 4.5000 & 9.0000 & 4.5000\\ 
4 &  9.4910 & 20.0000 & 4.7924 & 10.0000 & 4.0000\\ 
5 &  9.7815 & 21.2117 & 4.9835 & 10.6803 & 4.1666\\ 
6 &  9.9710 & 22.0078 & 5.1179 & 11.1724 & 4.0000\\ 
7 & 10.1034 & 22.5680 & 5.2176 & 11.5432 & 4.0888\\ 
8 & 10.2004 & 22.9811 & 5.2944 & 11.8326 & 4.0000\\ 
9 & 10.2742 & 23.2966 & 5.3553 & 12.0646 & 4.0500\\ 
10 & 10.3321 & 23.5446 & 5.4048 & 12.2545 & 4.0000\\ 
11 & 10.3784 & 23.7440 & 5.4458 & 12.4129 & 4.0333\\ 
12 & 10.4163 & 23.9074 & 5.4804 & 12.5469 & 4.0000\\ 
13 & 10.4478 & 24.0435 & 5.5099 & 12.6617 & 4.0238\\ 
14 & 10.4743 & 24.1583 & 5.5353 & 12.7612 & 4.0000\\ 
\hline
\rule[-3mm]{0mm}{8mm}
$\infty$ & 10.84 & 25.73 & 5.88 & 14.11 & 4.00\\ 
\hline
\end{tabular}
}
\caption{Eigenvalues  
in the adjoint case, the 't Hooft limit and the 
large $N_f$ limit. $B_1$ and $B_2$ are the lowest lying single
particle states in the $Z_2$ even and odd sectors, respectively.
The eigenvalue in the large $N_f$ limit is actually the continuum
threshold. Note that the masses are given in units $g^2N_c/\pi$ in
the first two cases and in units $g^2N_f/\pi$ in the third.}
\label{TableEigenvalues}
\vspace*{-0.4cm}
\end{table}

\subsection{Large $N_f$ limit and intermediate cases}
%\vspace{-0.1cm}

When going over to the large $N_f$ limit, $\lambda\rightarrow\infty$, 
the operator $P^-$, Eq.~(\ref{Pminus}), has a dominant kinetic term.
In the two-particle truncation the spectrum is purely kinematical. In the 
full theory a residual interaction is present.
In the spectrum,  Fig.~\ref{largeNSpectrum}(a), we see a continuum of states 
starting at $M^2=4g^2N_f/\pi$.
The only particle in the spectrum is thus a meson with
mass $M^2_M=g^2N_f/\pi$. This is in full agreement with results from analytic 
calculations \cite{AFST98}. The $Z_2$ odd sector looks 
similar, with the continuum starting at $M^2=9=3^2M^2_M$.
The single-particle state is absent in our calculations due to the 
tracelessness of the currents, {\em i.e.} the absence of the one current 
state.

Let us look at the eigenvalues as a function of $N_f/N_c$.  
Fig.~\ref{largeNSpectrum}(b) shows the first\footnote{M.~Engelhardt 
\cite{Engelhardt} presented a truncated version of this calculation.} 
calculation of the spectrum 
of two-dimensional Yang-Mills theories 
as a function of the continuous parameter $N_f/N_c$. 
We see that the (low-lying) trajectories are linear,
as is expected from the analytic calculations in 
Sec.~\ref{SecAnalytic}. 
The parameter $N_f$ actually plays a role similar to that of a mass term in 
the 't Hooft model \cite{Adi}.
%%%%%%%%%%%%%%%%%%%%%%%%%%%%%%% Changed 11/7/00 %%%%%%%%%%%%%%%%%%%%%%
Indeed, the curves obtained by taking the continuum limit 
of the single-particle trajectories in 
Fig.~\ref{largeNSpectrum}(b)
look very similar to results of massive two-dimensional QCD \cite{Peccei}. 
%%%%%%%%%%%%%%%%%%%%%%%%%%%%%%%%%%%%%%%%%%%%%%%%%%%%%%%%%%%%%%%%%%%%%%
%%%%%%%%%%%%%%%%%%%%%%%%%%%%%%% Changed 10/7/00 %%%%%%%%%%%%%%%%%%%%%%
Other DLCQ calculations of two-dimensional massive QCD
\cite{Hornbostel,Heyssler} were performed at finite $N_c$.
In particular, it was shown in Ref.~\cite{Hornbostel} that the spectrum 
of massive QCD in two dimensions 
is well described by the large $N_c$ approximation.
Deviations from the large $N_c$ behavior, found typically
at small fermion masses $m$ where the relevant coupling $g^2 N_c/m^2$ 
is strong, might just be artifacts of the DLCQ approach. 
This is in line with results of Ref.~\cite{AntonuccioPinsky98}, showing
that also in adjoint QCD in two dimensions 
the $1/N_c$ corrections are very small;
a fact that still awaits proper explanation. 
%%%%%%%%%%%%%%%%%%%%%%%%%%%%%%%%%%%%%%%%%%%%%%%%%%%%%%%%%%%%%%%%%%%%%%

We see a lot of level crossings in Fig.~\ref{largeNSpectrum}(b). 
The fact that the trajectories
do not intersect is a finite $K$ effect. We note that the plot shows a
smooth behavior at $\lambda=1$. There is no hint that the adjoint
case could be a special point in the parameter space.
%%%%%%%%%%%%%%%%%%%%%%%%%%%%%%%%%%%%%%%%%%%%%%%%%%%%%%%%%%%%%%%%%%%%%%
The eigenfunctions in the large $N_f$ limit 
look very different than those found in the 
adjoint and the 't Hooft case.
Typically, the amplitude vanishes, except for a few delta-like peaks,
reflecting the fact that the spectrum of this theory consists of
continuum states of non-interacting mesons.
%%%%%%%%%%%%%%%%%%%%%%%%%%%%%%%%%%%%%%%%%%%%%%%%%%%%%%%%%%%%%%%%%%%%%%
For comparison with other work, we list the eigenvalues obtained in
the numerical calculations in the 't Hooft limit, with adjoint fermions 
and in the large $N_f$ limit in Table \ref{TableEigenvalues}.

%----------------------------------------------------------------------%
%                       S U M M A R Y
%----------------------------------------------------------------------%
\vspace{-0.1cm}
\section{Summary and discussion}
%\vspace{-0.1cm}

%\begin{minipage}{15.556cm}
In this note we studied two-dimensional QCD using $SU(N_c)$
currents as basic degrees of freedom. 
This enabled us to calculate the Hamiltonian matrix and the spectrum of 
{any} two-dimensional $SU(N_c)$ gauge theory coupled to an arbitrary
number of fermions.
Working in the DLCQ framework, all relevant quantities become functions 
of the parameter $\lambda=N_f/N_c$ and of the harmonic 
resolution $K$. We constructed the Hamiltonian matrix explicitly in
the Fock basis of single-trace states. 
%This provides insight into the 
%differences of spectra as a function of $\lambda$. 

We reproduced all known results in the $\lambda$ parameter space: in 
the 't Hooft limit ($\lambda\rightarrow 0$) we obtained the well-known linear
spectrum, in the adjoint case
we find exactly the eigenvalues of previous works 
\cite{DalleyKlebanov93b,BhanotDemeterfiKlebanov93,GHK97},
and we identified the single meson of the large $N_f$ limit. 
Moreover, in the case of adjoint fermions we were able to confirm the 
single-particle nature of the lowest $Z_2$ odd boson, which was
hidden in Ref.~\cite{GHK97} among (trivial) multi-particle states which 
are absent in our calculation. 
We provided evidence for the single-particle nature of the 
boson $B_3$ at $M^2{=}45.99$, which
makes the existence of other single-particle states above 
the continuum threshold at $M^2=4M^2_{F_1}$ very likely.
We were able to explain the deformations in the trajectories of 
the eigenvalues as repulsions of eigenvalues. 
Unfortunately, the amount of remaining 'multi-particle' states in the 
spectrum makes it impossible to decide
whether or not the identified single-particle states form an infinite Regge
trajectory, or if even a multi-Regge structure exists, as was suggested in
Ref.~\cite{BoorsteinKutasov97}. 
What we were able to do is to eliminate all 
states from the list of single-particle candidates which do not appear
in our calculations, but in the bosonic sectors of previous works. 
By construction, our approach contains
{\em all} single-particle states \cite{KutasovSchwimmer95}. 

In Ref.~\cite{GHK97} it was suggested the massless limit is 
reached only in the
continuum limit $K\rightarrow\infty$. We work in a manifestly massless
approach and obtain exactly the same eigenvalues for all $K$. 
It would be very interesting to 
repeat the calculation of Ref.~\cite{GHK97} for finite $m$, as
was done with different motivation in Ref.~\cite{BhanotDemeterfiKlebanov93}.
With our results at hand, one can focus on a much smaller set of 
single-particle candidates, and see how these states develop as the
mass is turned on. Our guess is that this transition is continuous.
An interesting related question is how to distinguish screening from 
confinement when only the mass spectrum is known. 

We operate at higher numerical precision than previous work. This allowed us 
to prove numerically that the continuum threshold found in 
Ref.~\cite{GHK97} is indeed exactly at four times the mass squared of 
the lightest fermion. We note however that the interpretation of the
continuum states is not completely clear. We find that the deviation of
their masses from the expected free many-body masses, 
Eq.~(\ref{MultiParticleMass}), vanishes 
faster than $1/K^{2}$ as we go towards the continuum. If these states 
form a continuum at exactly the expected threshold, 
one has to conclude that they
decouple completely from the single-particle states and that their coupling 
is an artifact of the finite resolution $K$.

To summarize, we presented a refined and quite coherent picture of 
two-dimensional QCD by using a new computational tool. 
We hope that this approach 
will prove powerful and that it will yield new qualitative insight, too.
We mainly focused on quantitative improvements in the present work.
%However, we did not maximize our
%numerical efforts, because we might not ask the right 
%questions yet to completely understand the theory. 
We hope that this work will provide sufficient input for 
future enterprises to understand this theory, which shares some of the
key features of full QCD, much better.
%\end{minipage}

%----------------------------------------------------------------------%
%                       A C K N O W L E D G E M E N T S
%----------------------------------------------------------------------%

\vspace{0.4cm}
\centerline{\large\bf Acknowledgments}
\vspace{0.1cm}

The author thanks D.~Kutasov for initiating this work,
involvement in the analytic calculations and for many discussions, and 
is grateful for the hospitality during visits at the University of Chicago.
The author is grateful for many interesting discussions with A.~Armoni, 
Y.~Frishman and J.~Sonnenschein and also acknowledges discussions 
with F.~Antonuccio and S.~Pinsky.
A large part of this work was done while the author was 
supported by a Minerva Fellowship at the Weizmann Institute of Science,
Rehovot, Israel. 
The hospitality at the Weizmann Institute is gratefully acknowledged.
This work was supported in part by a Ohio State University 
Postdoctoral Fellowship.

%\vspace{-0.2cm}

%----------------------------------------------------------------------%
%                       B I B L I O G R A P H Y
%----------------------------------------------------------------------%

%\bibliography{KM_Paper_Biblio}

\begin{thebibliography}{99}

\bibitem{tHooft74b}
{G. 't Hooft},
%{\sl A two-dimensional model for mesons},
{\em Nucl.~Phys.~}{\bf B75} (1974) 461.
\bibitem{DalleyKlebanov93b}
	 {S.~Dalley, I.R.~Klebanov},
%	 {\sl String spectrum of 1+1 dimensional large N QCD with
%		  adjoint matter},
         {\em Phys.~Rev.~}{\bf D47}
	 (1993)
	 {2517--2527}.		 
\bibitem{Kutasov94}
	 {D.~Kutasov},
%	 {\sl Two dimensional QCD coupled to adjoint matter and
%		  string theory},
         {\em Nucl.~Phys.~}{\bf B414}
	 (1994)
	 33.
\bibitem{DateFrishmanSonnenschein87}
	 {G.D.~Date, Y.~Frishman, J.~Sonnenschein},
%	 {\sl The spectrum of multiflavor QCD in two dimensions},
	 {\em Nucl.~Phys.~}{\bf B283}
	 (1987)
	 {365--380}.
\bibitem{ArmoniSonnenschein95}
	 {A.~Armoni, J.~Sonnenschein},
%	 {\sl Mesonic Spectra of Bosonized $QCD_2$ Models},
	 {\em Nucl.~Phys.~}{\bf B457}
	 (1995)
	 81.
\bibitem{KutasovSchwimmer95}
	 {D.~Kutasov, A.~Schwimmer},
%	 {\sl Universality in two dimensional gauge theory},
         {\em Nucl.~Phys.~}{\bf B442}
	 (1995) 447.
\bibitem{AFST98}
	 {A.~Armoni, Y.~Frishman, J.~Sonnenschein, U.Trittmann},
%	 {\sl The spectrum of multi-flavor $QCD_2$ and the non-Abelian
%                  Schwinger equation},
	 {\em Nucl.~Phys.~}{\bf B537}	(1998)
	 {503--515}.
\bibitem{Smilga96}
	 {D.J.~Gross, I.R.~Klebanov, A.V.~Matytsin, A.V.~Smilga},
%	 {\sl Screening vs.~Confinement in 1+1 Dimensions},
	 {\em Nucl.~Phys.~}{\bf B461}
	 (1996)
	 109.
\bibitem{FrishmanSonnenschein95}
	 {Y.~Frishman and J.~Sonnenschein},
%	 {\sl $QCD_2$-screening, confinement and novel non-abelian
%		  solutions},
	 {\em Nucl.~Phys.~}{\bf B461}
	 (1997)
	 {285}; 	 
	 {A.~Armoni, J.~Sonnenschein},
%	 {\sl Screening and confinement in large $N_f$ $QCD_2$ and in 
%	 ${\cal N}=1$ $SYM_2$},
 	 {\em Nucl.~Phys.~}{\bf B502} (1997) 516--534.
\bibitem{ArmoniFrishmanSonnenschein98}
	 {A.~Armoni, Y.~Frishman and J.~Sonnenschein},
%	 {\sl The String Tension in Massive $QCD_2$},
	 {\em Phys.~Rev.~Lett.~}{\bf 80}
	 (1998)
	 430.
\bibitem{GHK97}
	 {D.J.~Gross, A.~Hashimoto, and I.R.~Klebanov},
%	 {\sl The spectrum of a Large $N$ Gauge Theory Near 
%                  Transition from Confinement to Screening},
	 {\em Phys.~Rev.~}{\bf D57}
	 (1998)
	 {6420--6428}.
\bibitem{AntonuccioPinsky98}
	 {F.~Antonuccio, S.S.~Pinsky},
%	 {\sl On the Transition from Confinement to Screening in
%                  $QCD_{1+1}$ Coupled to Adjoint Fermions at Finite $N$},
	 {\em Phys.~Lett.~}{\bf B439} (1998) 142--149.
\bibitem{BhanotDemeterfiKlebanov93}
	 {G.~Bhanot, K.~Demeterfi, and I.R.~Klebanov},
%	 {\sl 1+1 dimensional large N QCD coupled to adjoint fermions},
         {\em Phys.~Rev.~}{\bf D48}
	 (1993)
	 4980.
\bibitem{Bassetto98}
	 {A.~Bassetto},
%	 {\sl QCD: from four to two dimensions},
	 {\tt hep-th/9809084}.
\bibitem{PauliBrodsky85a}
	 {H.-C. Pauli, S.J. Brodsky}, 
%	 {\sl Solving field theory in one space and one time dimension},
	 {\em Phys.~Rev.~}{\bf D32}
         (1985)
         {1993--2000};
	%{\sl Discretized light-cone quantization: Solution to a 
        %       	  field theory in one space and one time dimension},
	 {\em Phys.~Rev.~}{\bf D32}
         (1985)
         {2001}.  
\bibitem{Witten84}
	 {E.~Witten},
%	 {\sl Non-abelian bosonization in two dimensions},
         {\em Comm.~Math.~Phys.~}{\bf 92}
	 (1984)
	 {455--472}.		  
\bibitem{Engelhardt}
	 {M.~Engelhardt},
%	 {\sl $QCD_{1+1}$ at large $N_f$ and $N_c$},
	 {\em Nucl.~Phys.~}{\bf B440},
	 (1995)
	 {543}.
\bibitem{Adi}
	{A.~Armoni},
	{\sl private communication}.
\bibitem{BoorsteinKutasov97}
	{J.~Boorstein, D.~Kutasov},
%	{\sl On the Transition from confinement to screening in
%		 large N gauge theory},
	in {Zinn-Justin} (Ed.), {\sl Les Houches 1997},
        Elsevier, Amster\-dam, 1997.
\bibitem{Peccei}
	 {A.J.~Hanson, R.D.~Peccei, M.K.~Prasad},
%	 {\em Two-dimensional SU(N) Gauge Theory, Strings and Wings:
%              comparative Analysis of Meson Spectra and Covariance},
         {\em Nucl.~Phys.~}{\bf B121} (1977) 477. 
\bibitem{Hornbostel}
	 {K.~Hornbostel, S.J.~Brodsky, H.-C.~Pauli}, 
%	 {\sl Light-cone quantized QCD in 1+1 dimensions},
	 {\em Phys.~Rev.~}{\bf D41}
         (1990)
         {3814--3821}.
\bibitem{Heyssler}
	 {M.~Heyssler, A.C.~Kalloniatis},
%	 {\sl Constituent Quark Picture out of QCD in two dimensions - 
%	      on the Light-Cone},
	 {\em Phys.~Lett.~}{\bf B354} (1995) 453--459.
\end{thebibliography}

%\end{document}

\end{document}